\documentclass[12pt,preprint]{aastex}

\newcommand{\vvm}{$V/V_{\rm max}$}
\newcommand{\vvmav}{$<V/V_{\rm max}>$}
\newcommand{\gz}{$GR(z)$}
\newcommand{\lufu}{luminosity function}
\newcommand{\ep}{$E_{\rm pk}$}
\newcommand{\epsp}{$E_{\rm pk}(sp)$}
\newcommand{\epob}{$E_{\rm pk,obs}$}
\newcommand{\epobvm}{$E_{\rm pk,obs}-V/V_{\rm max}$}
\newcommand{\liso}{$L_{\rm iso}$}
\newcommand{\epobliso}{$E_{\rm pk,obs}-L_{\rm iso}$}
\newcommand{\epliso}{$E_{\rm pk}-L_{\rm iso}$}

\newcommand{\epeiso}{$E_{\rm pk}-E_{\rm iso}$}
\newcommand{\sig}{$\sigma_{\log L}$}
\newcommand{\sw}{{\it Swift}}

\shorttitle{GRB Luminosity Function}
\shortauthors{Schmidt}

\begin{document}

\title{Gamma-Ray Burst Luminosity Functions Based On a Newly \\ 
Discovered Correlation Between Peak \\
Spectral Energy and \vvm.}

\author{Maarten Schmidt}
\affil{California Institute of Technology, Pasadena, CA 91125}

\begin{abstract}

We have discovered a correlation between the observed peak spectral 
energy \epob\ and the Euclidean value of \vvmav\ of gamma-ray bursts (GRBs). 
We present the evidence for the correlation in the GUSBAD catalog
and use it to derive the luminosity function of GRBs without using any
redshifts. The procedure involves dividing GUSBAD GRBs into five spectral
classes based on their \epob. The overall luminosity function is derived
assuming that each of the spectral classes contributes a gaussian
luminosity function. Their central luminosity is derived from the observed
Euclidean \vvmav. We explore various forms for the GRB rate function \gz\
in predicting redshift distributions of GRBs detected by \sw. We find
that \gz\ peaks at a higher redshift than the typical star formation 
history currently favored in the literature. We consider two examples
of \gz\ that sucessfully predict the observed redshift distribution of
\sw\ GRBs. With the luminosity functions in hand, we convert the \epobvm\ 
correlation into an \epobliso\ correlation and a rest frame \epliso\
correlation. In comparing the \epliso\ correlation with a published
correlation based on GRBs with known \epob\ and redshifts, we discuss 
the effect of Malmquist bias.

\end{abstract}


\section{Introduction}

The GRB \lufu\ plays a central role in the interpretation of the
observed source counts and redshift distributions of GRBs detected
with different instruments in different energy bands and to
different detection
limits. Of particular interest are the resulting rates and the
variation with redshift, since these must reflect the properties
of the GRB progenitors.

Ideally, the \lufu\ is derived from a well defined flux-limited 
sample of GRBs with redshifts. For most GRB surveys, it is not 
possible to collect such a sample with well defined gamma-ray and 
optical flux limits. We will discuss the situation for the \sw\ 
mission in Section 4.

In the absence of usable redshift statistics,
most studies of the \lufu\ use luminosity criteria, such as variability
\citep{fen00} and spectral lag \citep{nor00}. We have reviewed several
luminosity functions resulting from these studies and found that they
generally are not compatible with the source counts and the value of 
\vvmav\ for GRBs in the BATSE catalog \citep{sch04}. As long as this is
the case, they cannot provide reliable predictions about
redshift distributions, etc.

The correlation between the rest frame spectral peak energy \ep\
and the isotropic-equivalent radiated energy $E_{iso}$ found by 
\citet{ama02} seemed to be very promising as a luminosity indicator. 
However, it has been shown \citep{nak05,ban05} that for a large 
fraction of BATSE bursts there is no redshift that satisfies
the Amati relation; for further discussion also see \citet{ghi05}
and \citet{nak05}. According to \citet{lix07}, a redshift degeneracy 
in the Amati relation makes it impossible to derive redshifts larger 
than 0.9 with usable accuracy. Using durations and spectral parameters 
for GRBs detected by \sw\, \citet{but07} find a \epeiso\ correlation 
that is inconsistent with the Amati relation at $>5\sigma$ significance, 
with double the scatter. Both \citet{llo00} and \citet{but07} discussed
the effect of detector thresholds on the derivation of \epeiso\ 
correlations. The latter authors argue that all pre-\sw\ \epeiso\
correlations are likely unrelated to the physical properties of GRBs. 

In a study of the GRB luminosity function from 82 HETE-2 GRBs \citep{pel08},
an Amati-type correlation was used to derive pseudo-redshifts for the
62 GRBs lacking an observed redshift. This study provides information
about low-luminosity GRBs and X-ray flashes that are not represented in
the BATSE based GUSBAD catalog.

In the present study, we introduce a new correlation between the peak
spectral energy of the observed peak flux \epob\ and the Euclidean 
value of \vvmav. The correlation was discovered 
for GRBs in the GUSBAD catalog \footnotemark.
\footnotetext{Available at http//www.astro.caltech.edu/$\sim$mxs/grb/GUSBAD.} 
Since the Euclidean value of \vvmav\ is a geometric cosmological 
distance indicator \citep{sch01}, the peak luminosity of a subgroup of 
GRBs with given \epob\ can be derived without knowing any redshifts. 
Using the \epobvm\ correlation is not subject to any of the problems 
mentioned above for the Amati relation. 

The approach in our paper is the following. We use GUSBAD fluxes in 
the four BATSE DISCLA channels to divide them
into five spectral classes based on their \epob\ values.
We assume that the \lufu\ of each spectral group is a gaussian of
$\log L$ with a value of \sig\ large enough so that the overall sum of 
the five luminosity functions is reasonably smooth. In addition, we assume
that the overall rate density of GRBs varies with redshift as \gz.
The central luminosity $\log L_c$ of each spectral gaussian is derived 
by a process of iteration, until the corresponding value of \vvm\ for 
the spectral class agrees with that observed. 

The procedure allows deriving the overall GRB luminosity function
from the GUSBAD catalog without using any redshifts. The main free
parameter is the rate function \gz. To evaluate the predicted luminosity 
and redshift distributions for various forms of \gz, we do need a
sample with redshifts. We evaluate and use \sw\ detected GRBs
with observed redshifts. For the successful luminosity functions,
the process yields a rest frame \ep\ value for each of the spectral 
classes. This leads to \epobliso\ and \epliso\ correlations entirely 
based on data in this paper.

The use of \vvm\ and the
derivation of \epob\ from GRBs in the GUSBAD catalog are discussed in
Section 2, as well as the evidence for the \epobvm\ correlation. 
The framework for deriving the \lufu\ is discussed in Section 3. 
The status and completeness of \sw\ observations is covered in Section 4. 
In Section 5 we present two luminosity functions with differing redshift 
dependence \gz\ and test them on \sw\ redshift and luminosity distributions. 
The \epobliso\ and \epliso\ correlations are derived in Section 6.
A discussion and summary follows in Section 7.

\section{The \epobvm\ Correlation for GUSBAD GRBs}

\subsection{Euclidean Values of \vvm}

The GUSBAD catalog \citep{sch06} is based on observations with the 
BATSE LAD detectors which provide output in four energy channels, 
viz. $20-50$ keV (ch 1), $50-100$ kev (ch 2), $100-300$ keV (ch 3),
and $> 300$ keV (ch 4). The catalog lists peak photon fluxes for
channels 2 and 3 together. These were derived assuming a Band spectrum 
\citep{ban93} with $\alpha = -1.0$, $\beta = -2.0$, and $E_0 = 200$ keV. 
For the present study, we have derived for each GRB peak photon fluxes 
in each channel based on the two brightest illuminated LAD detectors. 

At the outset of this study, it was not clear whether the four channels
of the BATSE LAD detectors could be used for low-resolution 
spectrophotometry. We decided to use only GRBs with a 
$50-300$ keV peak photon flux $P$ exceeding 
0.50 ph cm$^{-2}$ s$^{-1}$, i.e. twice the effective photon flux limit 
of the GUSBAD catalog. This still leaves a large sample of 1319 GRBs. 
The Euclidean value of \vvm\ is simply \vvm\ $= (P/P_{lim})^{-3/2}$.

The value of \vvmav\ for uniformly distributed objects having any 
luminosity function in Euclidean space is $0.5$. This value derives 
from the fact that volumes are $\sim R^3$ and areas $\sim R^2$. 
This will not apply in non-Euclidean space, so for cosmological
objects the Euclidean value of \vvmav\ will deviate from $0.5$,
the more so the larger the typical redshift, i.e., it is a
cosmological distance indicator.

Concern has been expressed in the literature about the use of
\vvm\, particularly when sample thresholds vary 
\citep{ban92,pet93,har93}. The problem of varying threshold can be 
handled if each object has its own value of $P_{lim}$ as in the
GUSBAD catalog. In our present study, the adopted constant limit of
0.50 ph cm$^{-2}$ s$^{-1}$ is larger than all individual limits in
the catalog.

All well defined samples above a given flux limit are subject to
the Eddington effect \citep{edd13,edd40}. 
Random errors in the fluxes of individual
objects if positive may cause them to become part of the sample,
or, if negative, to be lost. Since there are generally more
objects in the flux distribution below any given flux than above,
the net effect will be that positive errors dominate in the sample. 
The effect on \vvmav\ will be that for objects with a uniform distribution, 
the observed value will be larger than 0.5.

We can evaluate the Eddington effect on \vvmav\ of our sample through
simulations. On the average, $\sigma$ for GUSBAD fluxes of the second 
brightest illuminated detector is $0.05$ ph cm$^{-2}$ s$^{-1}$.
Our adopted limit of $0.5$ ph cm$^{-2}$ s$^{-1}$ (for detector $2$)
corresponds to $\sim1.21$ ph cm$^{-2}$ s$^{-1}$ for detectors $1+2$, with
$\sigma = 0.071$ ph cm$^{-2}$ s$^{-1}$. Simulations show that in this
case the Eddington excess in \vvmav\ is $0.002$, considerably less
than the mean errors of \vvmav\ for the spectral samples discussed
below.

\subsection{Deriving Peak Energies}

Our goal is to derive the peak energy of the $\nu F_{\nu}$ spectrum.
Given the low spectral resolution, we initially derive 
$I(E_i) = E_i^2 N(E_i)$ for a representative energy $E_i$ in each 
BATSE channel $i$, where $N(E_i)$ is the photon flux density, in 
ph cm$^{-2}$ s$^{-1}$ keV$^{-1}$. 
We used the peak photon fluxes to derive the flux densities $N(E_i)$ 
at $E_i = $ 30, 70, 185 and 420 keV. For channels 1--3, these energies 
are close to the geometric means of the energy band limits; for channel
4, see below.

An initial estimate of the peak spectral energy is provided by the channel
with the largest value of $I(E_i)$. Table 1 shows for each of the four 
BATSE channels the number of GRBs having the peak at $E_i$. 
We can compare our \epob\ values with those given by \citet{kan06} for 
peak fluxes. Their work was based on BATSE LAD data of 350 bright GRBs 
in the BATSE catalogs. They used several LAD data types, and employed a
number of spectral shapes beyond the Band spectrum.
The Kaneko list has 219 GRBs that are in the GUSBAD catalog. The last
column of Table 1 gives the average Kaneko peak energy for the GUSBAD
sources in the Kaneko list. We adopt the Kaneko average for channel 4, 
since there is no well defined upper energy limit to its energy band.
Only for channel 3 is a meaningful comparison possible: our result
differs by only 10\% from the Kaneko average.  

The averages of the Euclidean values of \vvm\ for peak energies in 
channels 1-4 are 0.44, 0.45, 0.32, and 0.33, respectively. It turns 
out that the transition from high to low values of \vvmav\ takes 
place within channel 3. Since it contains a large number of peak 
energies, we subdivide those in channel 3 based on whether the ratio 
of $I(E_4)/I(E_2)$ is (a) less than 0.5, (b) in the range $0.5 - 1.0$, 
or (c) larger than 1.0. Given that only 24 objects peak in channel 1,
we combine channels 1 and 2. We end up with five spectral peak classes
$sp$ with \vvmav\ ranging from $\sim 0.45$ to $\sim 0.30$, see Table 2. 
The resulting \epobvm\ correlation is illustrated in  Figure 1. 
 
Table 2 also gives the mean values of $\alpha_{23}$, the photon spectral 
index derived from BATSE channels 2 and 3, as given in the GUSBAD catalog.
The $\alpha_{23}-V/V_{\rm max}$ correlation had been noted before
\citep{sch01} in a study similar to the present one.

\section{Derivation of Source Counts from the Luminosity Function}

Since we do not have redshifts for our sample of GUSBAD GRBs, we
cannot derive the luminosity function directly. Instead we iterate
the luminosity function by trial and error. In each step of the
iteration we derive the predicted \vvmav\ value for each of the 
spectral classes until they agree with those given in Table 1.

We assume that the GRB luminosity function $\Phi(L,z,sp)$ of 
spectral class $sp$ can be written as
\begin{equation}
 \Phi(L,z,sp) = \Phi_0(L,sp) GR(z),
\end{equation}
where $L$ is the peak luminosity, $\Phi_0(L,sp)$ is the $z=0$ 
luminosity function of class $sp$, and \gz\
the comoving GRB rate density, normalized at $z=0$.
We assume that $\Phi_0(L,sp)$ has a gaussian distribution 
\begin{equation}
 \Phi_0(L,sp) = {R_0(sp) \over {\sigma_{\log L} \sqrt(2\pi)}} 
  \exp{-[\log L - \log L_c(sp)]^2 \over 2 \sigma_{\log L}^2}
\end{equation}
with dispersion $\sigma_{\log L}$ around a central peak 
luminosity $L_c(sp)$. $R_0(sp)$ is the GRB rate at $z=0$.

Our approach will be to iterate the value of $L_c(sp)$ until the
corresponding value of \vvmav$(sp)$ for the GUSBAD sample agrees with 
the observed value. This requires a full derivation of the 
expected source counts 
from the \lufu. The role of $\sigma_{\log L}$ is primarily to make each
spectral \lufu\ sufficiently wide so that the total \lufu\ is reasonably
smooth. We ended up using $\sigma_{\log L} = 0.5$.

In deriving fluxes from luminosities and redshifts, we employ the BATSE 
energy range $(E_1,E_2)$ with $E_1 = 50$ keV and $E_2 = 300$ keV. 
For an object at redshift $z$, the observed energy range $(E_1,E_2)$
originates in the range $(E_1(1+z),E_2(1+z))$ in the object's
rest frame, whereas the luminosity refers to the range $(E_1,E_2)$.  
The K-term is the ratio of the rest frame energies radiated 
in the two ranges,
\begin{equation}
K(z) = {{{\int^{E_2(1+z)}_{E1(1+z)} E N(E,E_{pk},\alpha,\beta)} dE} 
\over {{\int^{E_2}_{E_1} E N(E,E_{pk},\alpha,\beta)}} dE} .
\end{equation}
The Band photon spectrum is ususally described in terms of a break
energy $E_0$. Here we use a Band spectrum $N(E,E_{pk},\alpha,\beta)$ where
$E_{pk}(sp)=(2 + \alpha)E_0(sp)$ assuming that $\beta < -2$ \citep{ban93}. 
We adopt constant values of $\alpha = -0.8$ and $\beta = -2.6$ for 
reasons that will be discussed in Section 5. 

The peak flux $P(L,z)$ observed for a GRB of luminosity $L$ at redshift
$z$ is
\begin{equation}
P(L,z) = {L\over 4\pi ((c/H_0)A(z))^2}{K(z)},
\end{equation}
where $(c/H_0)A(z)$ is the bolometric luminosity distance.
We use the cosmological parameters $H_0 = 70~$km s$^{-1}$ Mpc$^{-1}$,
$\Omega_M = 0.3$, and $\Omega_{\Lambda} = 0.7$. 

The integral peak flux distribution for GRBs of spectral class $sp$ is, 
\begin{equation}
N(>P,sp) = \int \Phi_o(L,sp)\,dL  
\int^{z(L,P,sp)}_0 GR(z)(1+z)^{-1}\, (dV(z)/dz)\, dz,
\end{equation}
where $z(L,P,sp)$ is derived from equation (4), $V(z)$ is the
comoving volume and the term $(1+z)^{-1}$ represents the time
dilation. With this formulation, it is straightforward to derive
the differential source counts $dN(>P,sp)/dP$, as well as the average
values of \vvm, \epob, $\alpha_{23}$, etc. 

For a given rate function \gz\, the procedure to iterate the luminoxity
function is as follows. Assume starting values for the central 
luminosity $L_c(sp)$ and the rest frame Band peak energy $E_{pk}(sp)$.
The differential source counts together with 
\vvm\ $= (P/P_{lim})^{-3/2}$ produce the expected values of \vvmav$(sp)$ 
for the GUSBAD catalog. Similarly, the expected values of \epob$(sp)$ are 
obtained by weighting \epsp$/(1+z)$ with the differential
source counts. The iteration is repeated until the expected values of
\vvmav$(sp)$ and \epob$(sp)$ match the observed ones given in Table 2. 

It is worth noting that given \gz\ and the shape of the five spectral
luminosity functions, the procedure leads for each $sp$ to a single 
value for $L_c(sp)$ and $E_{pk}(sp)$. The primary unknown is the 
density function \gz. We will use \sw\ data to test various forms 
of \gz, see Sec. 5.

\section{\sw\ Data}

The \sw\ mission \citep{geh04} is in the process of detecting 
hundreds of GRBs. Thanks to an emphasis on rapid identification and
communication, relatively many of these GRBs have observed redshifts.
We use this data base to check predicted distributions of redshifts
and luminosities based on luminosity functions with different \gz. 
 
We have used \sw\ GRB data for the period Jan 1, 2005 -- Sep 30, 2008. 
The observations cover a field of view of 1.4 sr in the energy band 
$15-150$ keV. We use observed peak fluxes over a one second time interval 
obtained from the High Energy Astrophysics Science Archive Research Center 
(HEASARC) and redshift information from J. Greiner's list \footnotemark.
\footnotetext{Available at http://www.mpe.mpg.de/$\sim$jcg/grbgen.html.} 
We derive isotropic-equivalent luminosities for the energy band $15-150$
keV using the rest frame \epliso\ correlation given in Section 6. 
In deriving these luminosities, we assumed a Malmquist correction
$\Delta \log L = +0.36$, see Section 6.

The \sw\ mission does not provide a flux limit above which 
the sample of GRBs is complete. \citet{ban06} has analyzed the sensitivity
of \sw\ in detail. He concludes that the complexity of the trigger
system maximizes the sensitivity, but ``makes an accurate determination
of this sensitivity at a given time very difficult if not impossible.''
Given these circumstances, we have to carry out an {\it a posteriori} 
estimate of the effective flux limit above which the burst list is complete. 

Following the procedure outlined in Section 3, we can derive the predicted
distributions of fluxes, luminosities and redshifts for \sw. The
predicted flux distribution turns out to be virtually the same for
all forms of \gz\ discussed in the next section. We show the
distribution in Figure 2, together with the observed number of GRBs
and the number of GRBs with observed redshifts. At large fluxes, the 
prediction appears to be around $9\%$ below the observed numbers.
The observed numbers start to deviate systematically
below the prediction at $P = 1.0$ ph cm$^{-2}$ s$^{-1}$. We adopt
this flux as the effective completeness limit for the \sw\ bursts. 
The total number of GRBs above the flux limit is 217, of which 84 
have redshifts. The fraction of GRBs with redshifts declines from 
$\sim 0.6$ at large flux to $\sim 0.35$ at the limit. 
We will assume that a redshift fraction of 0.4 applies uniformly above 
the flux limit. 

With the \sw\ flux limit and the fraction with redshifts set, we are
now in a position to derive from the luminosity function for a given \gz\
the predicted distributions of luminosity and redshift for the \sw\
sample.

\section{Exploring Luminosity Functions with Different \gz\ Functions}

As discussed at the end of Sec. 3, the only unknown in 
deriving the GRB luminosity function, given the \epobvm correlation, 
is the redshift dependence \gz. Therefore, we can view the excercise
as one in which we explore which shape of \gz\ is compatible with the
observations based on \sw\ data.

We first explore the case where \gz\ is the galaxy star formation 
history (SFH). In the many published studies of star formation in the 
literature, the SFH rises fast with redshift by about an order 
of magnitude out to $z \sim 1.5$, beyond which it levels off; it tends 
to decline beyond $z \sim 3$. We use the analytical expression, 
normalized to $z=0$,
\begin{equation}
 \rho_{SFH} = (1.0 + (0.10/0.015)z)/(1.0 + (z/3.4)^{5.5}),
\end{equation}
for the SFH described by \citet{hop06}, based on an extensive 
compilation by \citet{hop04}.  

The predicted distributions of luminosity and redshift are compared
to the \sw\ observations in Figures 3 and 4. The predicted number of 
GRBs (with redshifts) is 79.6, while the observed number is 84. 
The prediction is $\sim 5\%$ below the observed number. 
It is obvious that the agreement with the observations is poor, both
for the luminosities and the redshifts. We have carried out Monte Carlo 
simulations to find the probability $P(\log L > 51.5)$ that the observed 
number of luminosities above $\log L = 51.5$ can be produced from the
predicted distribution by chance. We find $P(\log L > 51.5) < 10^{-6}$. 
Similarly, for redshifts above 3.0 the probability
$P(z > 3.0) < 10^{-6}$, see Table 3.
Even if we continue the SFH beyond the peak redshift $z = 2.55$ at its
peak value without any downturn, the agreement remains poor, with
$P(\log L > 51.5) < 10^{-6}$ and $P(z > 3.0) = 0.007$. Clearly, 
the SFH cannot represent \gz.  

With little guidance as to what the shape of \gz\ could be, we use a 
simple schematic involving a $(1+z)$ power law rise, a plateau, and
an exponential decline with $z$, as follows:
\begin{eqnarray}
 & & GR(z)=(1+z)^m \;\;\;\;\;\;\;\;\;\;\;\;\;\;\;\;\;\;\;\; 0<z<z_c\\
\label{eq:gz1}
 & & GR(z)=(1+z_c)^m \;\;\;\;\;\;\;\;\;\;\;\;\;\;\;\;\;\; z_c<z<z_d\\
\label{eq:gz2}
 & & GR(z) = (1+z_c)^m 10^{k(z-z_d)} \; \; \; \; \; z > z_d
\label{eq:gz3}
\end{eqnarray}  

With the goal of producing redshift and luminosity distributions
more in accord with the observed \sw\ distributions, we explored a
number of combinations of the free parameters $m$ and $z_c$. 
We concluded from these explorations that to first order
the ratio $R=GR(z=4)/GR(z=1.5)$ is crucial. This appears related
to the fact that most \sw\ bursts have redshifts below 2.
For the SFH considered above, the ratio $R=0.74$. 

We present detailed results for two shapes of \gz\ that 
have R-values of 3.2 (A) and 4.0 (B), respectively. Table 4 and 
Figure 5 illustrate the two cases. Case A rises by a factor of 32
to $z=3.0$ and then remains constant. Case B rises slower to a
plateau of 25 starting at $z=4.0$. We have not actually included 
the exponential decline beyond $z_d$. With ony two redshifts with
$z>5.0$ in the \sw\ sample, we have no observational leverage
on the value of $z_d$.

The properties of the two models are given in Tables 3 and 4.
Both are successful in producing luminosity and redshift 
distributions compatible with the \sw\ data, see Figures 6 and 7.
The redshift distributions predict that $3-4\%$ of GRBs in
the \sw\ sample have $z>6$ and $\sim 1\%$ $z>8$. Since we have
allowed the density plateau to continue beyond $z=10$,
these percentages are overestimates. Our predictions are 
lower than most discussed in the literature.

In deriving the luminosity functions, we found that the spectral 
indices $\alpha_{23}$ given in Table 2 are best represented by the 
Band spectral parameters $\alpha = -0.8$ and $\beta = -2.6$. We show 
the observed average spectral index $\alpha_{23}$ for the five 
spectral classes versus \epob\ in Figure 8, together with the
relation derived from the luminosity functions.
 
We illustrate in Figure 9 the derivation of $L_c(sp)$ from the
observed \vvmav\ values for case A. The differences in the curves 
reflect the different values of \ep\ for the spectral classes. 
For the GUSBAD sample, the curves are only meaningful near the
actual $L_c(sp)$ values. For, say, a deeper sample, the applicable
parts would be at correspondingly higher luminosities.

Figure 10 shows the predicted distribution of redshifts for
the GUSBAD sample, again based on case A. For $sp = 1$ essentially
all the redshifts are expected to be below 1. This component is very 
uncertain, as discussed below. The peak redshifts for the different
spectral classes reflect their \vvmav\ values. This illustrates clearly
that the Euclidean \vvmav\ is a cosmological distance indicator.

In Figure 11 we show the luminosity function for case A, as
well as the luminosity distribution for the 1319 GUSBAD sources
with $P > 0.5$ ph cm$^{-2}$ s$^{-1}$. Also shown are the individual
luminosity functions for the five spectral classes. The first peak 
of the luminosity function is contributed by spectral class 1. 
The lower half of its gaussian clearly plays no role, as it produces
no objects in the luminosity distribution. The luminosity assigned 
to this class is uncertain, since the slope of the curve in
Figure 9 is relatively shallow. Actually, if the \vvmav\ for $sp=1$
were only $1.2\sigma$ larger, it could not be reproduced by any value
of $L_c$. Altogether, this suggests that the 
(large) $z=0$ density rates $R_0$ for $sp = 1$ given in Table 4 
are very uncertain. The second peak in the luminosity function is 
contributed by the large number of GRBs in spectral classes 
2-5. Their combined $z=0$ rate is $0.09 - 0.22$ Gpc$^{-3}$ y$^{-1}$,
for models A and B, respectively.

\section{The \epobliso\ and \epliso\ correlations}

\ep\ correlations with radiated energy or luminosity are of interest
in exploring the mechanism for the prompt emission of GRBs. They
are also of practical interest in allowing an estimation of the
redshift of GRBs with measured \epob. In this section, we discuss
the derivation of the relevant isotropic-equivalent luminosity \liso, 
present the \epobliso\ and \epliso\ correlations and briefly mention 
the problem of extracting individual redshifts from \ep\ correlations.

As discussed in Section 3, the derivation of the luminosity function
involves an iteration of the central luminosity $L_c(sp)$, where
each spectral component is a gaussian with dispersion \sig.
We chose a dispersion \sig $= 0.5$ that produces a reasonably smooth 
overall luminosity function. We want to use the $L_c(sp)$ luminosities
in deriving the isotropic-equivalent peak luminosities \liso\ used 
in the correlations. 

It turns out, however, that $L_c(sp)$ varies considerably with \sig.
We explored using \sig\ values ranging from 0.1 to 0.6 in deriving
the \lufu\ for cases A and B. For all values of \sig\ 
the agreement with the observed luminosities and redshifts
of the \sw\ GRBs is good. The variation of $\log L_c(sp)$ appears to 
be proportional to $\sigma_{\log L}^2$.  

The explanation of the situation is as follows. As \sig\ is increased,
the wings of the gaussian will provide more objects with higher and
lower luminosities. The higher luminosities will be at larger
distances and contribute lower \vvm\ values. The lower luminosities
will contribute higher \vvm\, but the shift will be less due to the
curvature of the relation between $L_c(sp)$ and \vvm, see Figure 9.
The asymmetry results in a
higher value of \vvmav\, requiring a shift to a lower $L_c(sp)$ 
in order to fit the observed input value of \vvmav. The reason why
agreement with the \sw\ observations is not affected, is that the
additional objects of lower luminosity contribute little to the
observed distributions because they are observed over much smaller
volumes.

Given our understanding of the variation of $L_c(sp)$ with \sig,
it is clear that we want to use the unbiased values of $\log L_c(sp)$
at \sig $= 0$ in deriving the \liso. They are only $0.01-0.03$ 
larger than the $\log L_c(sp)$ values for \sig $= 0.1$. The \liso\ 
values given in Table 4 are $2-5$ times larger than $L_c(sp)$.

We now plot the \liso\ versus \epob\ to produce the \epobliso\
correlation in Figure 12. The error bars for \liso\
reflect the effect of those in \vvmav. The line
drawn represents a regression of \liso\ on \epob, assuming that
the \epob\ values are errorless,
\begin{equation}
\log L_{\rm iso} = 51.0+2.52^{+0.37}_{-0.41}(\log E_{\rm pk,obs}
 - 2.08^{+0.39}_{-0.36}).
\end{equation}

The derivation of the luminosity function involved an iteration 
not only of the central luminosity $L_c(sp)$, but also of the
rest frame value of the Band peak energy $E_{\rm pk}(sp)$.
Figure 13 shows the \epliso\ correlation.
The line drawn represents a regression of \liso\ on \ep,
\begin{equation}
\log L_{\rm iso} = 51.0 + 1.75^{+0.26}_{-0.28}(\log E_{\rm pk}
 - 2.36^{+0.46}_{-0.42}).
\end{equation}
The extraction of individual redshifts from the \epobliso\ correlation 
for sources with known \epob\ is straightforward. Using the
discrete \epob\ values of Table 2, we can produce redshifts for all 
1319 GRBs without problems. Since the correlation uses the \epob\
in the observer's frame, it can only be used for GRBs with a 
$50-300$ keV flux larger than 0.5 ph cm$^{-2}$ s$^{-1}$.
 
The situation is entirely different for the \epliso\ correlation. 
When we use the correlation to derive redshifts for sources with
known \epob, we find that only 772 out of 1319 GRBs, or 59\%, yield 
a redshift. The situation is strikingly similar to that reported for 
the Amati relation \citep{ama02} by \citet{nak05} and \citet{ban05},
who found that for a large fraction of BATSE bursts there is no redshift 
that satifies the relation.

There are relatively few \epliso\ correlations reported in the litterature 
\citep{scf07,yon04}. We will compare our correlation with the one given
by Schaefer, based on 62 GRBs with \epob\ values and redshifts,
\begin{equation}
\log L(\rm bol)_{\rm iso} = 52.21 + 1.68 (\log E_{\rm pk}/ 300 {\rm keV}). 
\end{equation}
Here $\log L(\rm bol)$ covers the energy band $1-10000$ keV. Transformation
of our $50-300$ keV \liso\ values listed in Table 4 to $1-10000$ keV
increases them on the average by $0.61$ in $\log L$. Comparison of the
bolometric luminosities yields 
\begin{equation}
<\log L(\rm Schaefer) - \log L(\rm present)> = +0.52 \pm 0.22. 
\end{equation}

Next we consider the effect of Malmquist bias \citep{mal36}, the 
difference in average luminosity of objects observed in a sample 
above a given flux limit and the average luminosity in space, 
\begin{equation}
\Delta \log L = <\log L>_{obs} - <\log L>_{space}.
\end{equation}
In Euclidean space, the bias is proportional to $\sigma_{\log L}^2$. 
For cosmological objects with varying co-moving density, the situation
is more complex. 

All studies of individual GRBs with redshifts aimed at 
deriving \ep\ correlations are subject to Malmquist bias since they can 
only be carried out above some flux limit. The Malmquist bias depends 
on the luminosity function (in space) of GRBs at given \ep. 
This function is not known at present.

We do have clear evidence of Malmquist bias in the present study.
Figure 11 shows that compared
to the luminosity functions for the five spectral classes, 
the luminosity distributions predicted for the GUSBAD sample are shifted 
to considerably higher luminosities. The Malmquist biases for $sp = 1-5$ 
are $0.70, 0.53, 0.31, 0.28$ and $ 0.36$, respectively, for
an average of $0.44$. This value is tantalizingly close to the offset
shown in eq. (13). However, our values are based on an assumed 
gaussian shape of the spectral luminosity functions with a dispersion
of $\sigma_{\log L} = 0.5$, designed to make the overall luminosity
function reasonably smooth. As long as we do not have reliable information
about the distribution of $\log L_{\rm iso}$ at given \ep\ in space,
we will not be able to assess the effect of Malmquist bias accurately. 
 
\section{Discussion}

We present in this final section some commentary on the approach 
used in this paper and on the results.

1. This work validates the use of \vvmav\ as a cosmological distance
indicator. It requires a complete sample of GRBs above a well defined
flux limit. The GUSBAD catalog with its uniform treatment of all
sources is eminently useful for this purpose. The input in the form 
of \vvmav\ forced us to a deductive approach where we used a given shape 
of the (spectral) luminosity function and iterated its center luminosity 
and rest frame \ep\ until the observed \vvmav\ and \epob\ were fit. 
An important advantage of the deductive method is that the derivation 
of source counts from a given luminosity function can be carried out 
accurately.

2. This approach is in contrast to the inductive method generally used 
when redshifts are available. This involves deriving densities in 
bins of redshift and luminosity and correlating these 
with redshift and/or luminosity to derive the \lufu. In such an
approach, it is essential to check that the resulting luminosity 
function produces the input source counts and \vvmav\ correctly.
If this is not the case, predictions about expected numbers of
sources at large redshift will be unrealistic \citep{sch04}.

3. If GRBs were not subject to cosmological evolution, the input 
in terms of \vvmav, \epob, and $<\alpha_{23}>$ would suffice to 
derive the \lufu\ and hence predict source counts and the redshift 
and luminosity distributions. Comparing these predictions with the 
observations, such as from \sw, would then constitute a check on 
the cosmological model. In reality, there is evolution, so in the 
context of the present study the \sw\ observations of GRBs with 
redshifts essentially provide information about \gz.

4. We have indicated that our exploration of various forms for \gz\
indicate that the \sw\ observations seemed to require that the 
$GR(z=4)/GR(z=1.5)$ ratio be at least 3. None of the star formation 
results shown in Figure 1 of \citet{hop06} exhibit such a ratio. 
\citet{le06} explored a variety of GRB rate curves and
found the best fit for their model SFR6, for which 
$GR(z=4)/GR(z=1.5) \sim 2$. Given that GRBs tend to occur in blue,
underluminous galaxies of limited chemical evolution \citep{fru06},
it is not surprising that \gz\ is different from the main cosmic star
formation, and concentrated to earlier cosmic times.
A study of the hydrogen ionization rate, based on the Lyman-$\alpha$ 
forest \citep{fau08}, concludes that the star formation rate is 
increasing for $z = 2 - 4$, qualitatively much like our model B.

5. We have been conservative in setting the effective completeness
limit for the \sw\ data at $P = 1.0$ ph cm$^{-2}$ s$^{-1}$ and 
using only data above the limit. It would be helpful if catalogs
of \sw\ sources could include an indication for each source whether
it can be considered to be part of a sample that is complete above a 
given photon limit over a given area.

6. Even while not all individual GRBs can be fitted with a Band spectrum,
it appears that for statistical work, the \epliso\ correlation together 
with fixed values for $\alpha = -0.8$ and $\beta = -2.6$ provide a
surprisingly simple prescription.

7. In comparing \epliso\ and \epeiso\ correlations, the transformation
from the actual energy bands used, like $50-300$ or $15-150$ keV, to 
$1-10000$ keV \citep{ama02} introduces needless uncertainty. It would
be preferable to report results in the observed energy bands. 

\acknowledgments

This research made use of data obtained from HEASARC, provided by 
NASA's Goddard Space Flight Center. It is a pleasure to thank 
Y. Kaneko for detailed information about spectra.

\clearpage

\begin{deluxetable}{ccccc}
\tablecaption{Spectral Peak Energies\label{tbl-1}}
\tablewidth{0pt}
\tablehead{
\colhead{Channel} & \colhead{n\tablenotemark{a}} & 
\colhead{\epob (keV)\tablenotemark{a}} & 
\colhead{n\tablenotemark{b}} & 
\colhead{\epob (keV)\tablenotemark{b}} 
}
\startdata
 1 &   24  &  30   &   ..   &  ..  \\
 2 &  204  &  70   &    4   &  75  \\ 
 3 &  608  & 185   &   99   & 202  \\ 
 4 &  483  & 420\tablenotemark{c}   &  116   & 417 \\
\enddata
\tablenotetext{a}{GUSBAD sample}
\tablenotetext{b}{\citet{kan06} data}
\tablenotetext{c}{Value adopted from \citet{kan06}}
\end{deluxetable}

\clearpage

\begin{deluxetable}{cccccc}
\tablecaption{Peak Spectral Energy, \vvmav\ and Photon Index \label{tbl-2}}
\tablewidth{0pt}
\tablehead{
\colhead{Spec. Cl.} &  \colhead{Channel} & \colhead{n} & 
\colhead{\epob (keV)} & \colhead{\vvmav} & \colhead{$<\alpha_{23}>$} 
}
\startdata
 1 & 1,2 &                228  &  65   &  0.449$\pm$0.019   &  -2.52  \\ 
 2 & 3\tablenotemark{a} & 185  & 120   &  0.395$\pm$0.021   &  -1.77  \\ 
 3 & 3\tablenotemark{b} & 207  & 175   &  0.296$\pm$0.018   &  -1.75  \\
 4 & 3\tablenotemark{c} & 216  & 250   &  0.283$\pm$0.018   &  -1.44  \\
 5 & 4 &                  483  & 420   &  0.327$\pm$0.013   &  -1.25  \\
\enddata
\tablenotetext{a}{$I(E_4)/I(E_2) < 0.5$}
\tablenotetext{b}{$0.5 < I(E_4)/I(E_2) < 1.0$}
\tablenotetext{c}{$1.0 < I(E_4)/I(E_2)$}
\end{deluxetable}

\clearpage

\begin{deluxetable}{cccc}
\tablecaption{Comparison of Luminosity Function Models \label{tbl-3}}
\tablewidth{0pt}
\tablehead{
\colhead{Model} &  \colhead{SFH} & \colhead{A} & \colhead{B} 
}
\startdata
 \gz\ & $\rho_{SFH}$\tablenotemark{a} & $(1+z)^{2.5}$ & $(1+z)^{2.0}$ \\
 $z_c$       &              &    3.0    &   4.0 \\ 
 $z_d$       &              &   10.0    &  10.0 \\
 $k$         &              &    0.0    &   0.0 \\
 $P(\log L > 51.5)\tablenotemark{b}$ &  $< 10^{-6}$ & 0.16 &  0.18\\
 $P(z > 3.0)\tablenotemark{c}$       & $< 10^{-6}$  & 0.93 &  0.97 \\
 $f(z>2)$\tablenotemark{d} & 0.198 & 0.448 & 0.426  \\
 $f(z>4)$\tablenotemark{d} & 0.009 & 0.126 & 0.159  \\
 $f(z>6)$\tablenotemark{d} & 0.000 & 0.037 & 0.049  \\
 $f(z>8)$\tablenotemark{d} & 0.000 & 0.012 & 0.017  \\
 
\enddata
\tablenotetext{a}{Analytical expression for the star formation
history SFH \citep{hop06}, see eq. (6)}
\tablenotetext{b}{Probability that the model produces the observed number
of \sw\ bursts with $\log L > 51.5$}
\tablenotetext{c}{Probability that the model produces the observed number
of \sw\ bursts with $z>3$}
\tablenotetext{d}{Fraction of \sw\ bursts with 
$P > 1.0$ ph cm$^{-2}$ s$^{-1}$ above the given redshift}

\end{deluxetable}

\clearpage

\begin{deluxetable}{cccrlcccrl}
\tablecolumns{10}
\tablewidth{0pc}
\tablecaption{Spectral Class Luminosity Functions \label{tbl-4}}
\tablehead{
\colhead{}    &  \multicolumn{4}{c}{Case A} &   \colhead{}   &
\multicolumn{4}{c}{Case B} \\
\cline{2-5} \cline{7-10} \\
\colhead{Spec. Cl.} & \colhead{$\log L_c$\tablenotemark{a}} &
\colhead{$\log L_{\rm iso}$\tablenotemark{b}} & 
\colhead{$E_0$\tablenotemark{c}} & 
\colhead{$R_0$\tablenotemark{d}} & \colhead{} & 
\colhead{$\log L_c$\tablenotemark{a}} & 
\colhead{$\log L_{\rm iso}$\tablenotemark{b}} & 
\colhead{$E_0$\tablenotemark{c}} & 
\colhead{$R_0$\tablenotemark{d}}}
\startdata
1 & 49.11$^{+0.67}_{-0.57}$& 49.81$^{+0.54}_{-0.33}$
&   73 & 6.8\tablenotemark{e} &   &  48.93$^{+0.45}_{-0.67}$
&   49.63$^{+0.58}_{-0.47}$ &   69 & 13.0\tablenotemark{e}\\ 
2 & 50.84$^{+0.27}_{-0.38}$ & 51.53$^{+0.23}_{-0.33}$ &  240 & 0.048  &   
&   50.54$^{+0.36}_{-0.44}$ & 51.14$^{+0.31}_{-0.44}$ &  203 & 0.149  \\ 
3 & 51.67$^{+0.12}_{-0.12}$ & 51.97$^{+0.06}_{-0.06}$ &  506 & 0.0089 &   
&   51.71$^{+0.13}_{-0.15}$ & 52.07$^{+0.07}_{-0.07}$ &  518 & 0.0133 \\ 
4 & 51.70$^{+0.11}_{-0.12}$ & 51.95$^{+0.06}_{-0.06}$ &  762 & 0.0078 &   
 &  51.75$^{+0.12}_{-0.13}$ & 52.05$^{+0.06}_{-0.06}$ &  792 & 0.0112 \\ 
5 & 51.39$^{+0.08}_{-0.08}$ & 51.76$^{+0.04}_{-0.04}$ & 1217 & 0.0225 &   
&   51.42$^{+0.08}_{-0.09}$ & 51.85$^{+0.04}_{-0.04}$ & 1244 & 0.0336 \\ 
\enddata
\tablenotetext{a}{Central isotropic-equivalent luminosity of the gaussian 
in the $50-300$ keV energy band for \sig $ = 0.5$, in erg s$^{-1}$.
Errors correspond to those for \vvmav, see Table 2.} 
\tablenotetext{b}{Isotropic-equivalent peak luminosity in the $50-300$ keV 
energy band, in erg s$^{-1}$, see Sec. 6. 
Errors correspond to those for \vvmav, see Table 2.} 
\tablenotetext{c}{Band spectrum break energy in the rest frame, in keV}
\tablenotetext{d}{GRB density rate at $z=0$, in Gpc$^{-3}$ y$^{-1}$} 
\tablenotetext{e}{Rates for $sp=1$ are very uncertain, see Sec. 5.}

\end{deluxetable}

\clearpage

\begin{figure}
\includegraphics[angle=270,scale=.80]{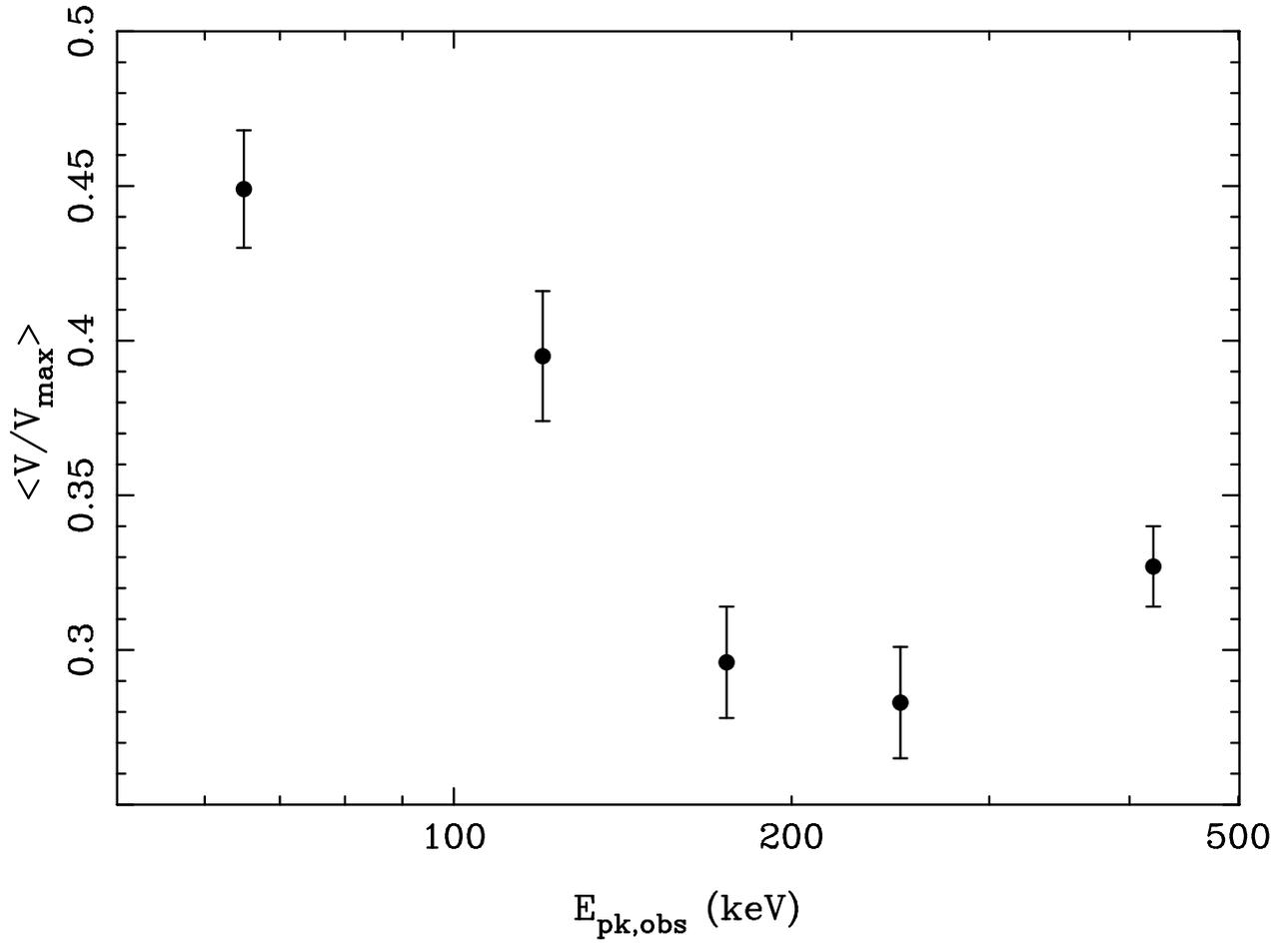}
\caption{\vvmav\ values versus the observed peak spectral 
energy \epob\, based on 1319 GUSBAD GRBs with 
$P > 0.5$ ph cm$^{-2}$ s$^{-1}$.\label{fig1}}
\end{figure}

\begin{figure}
\includegraphics[angle=270,scale=.80]{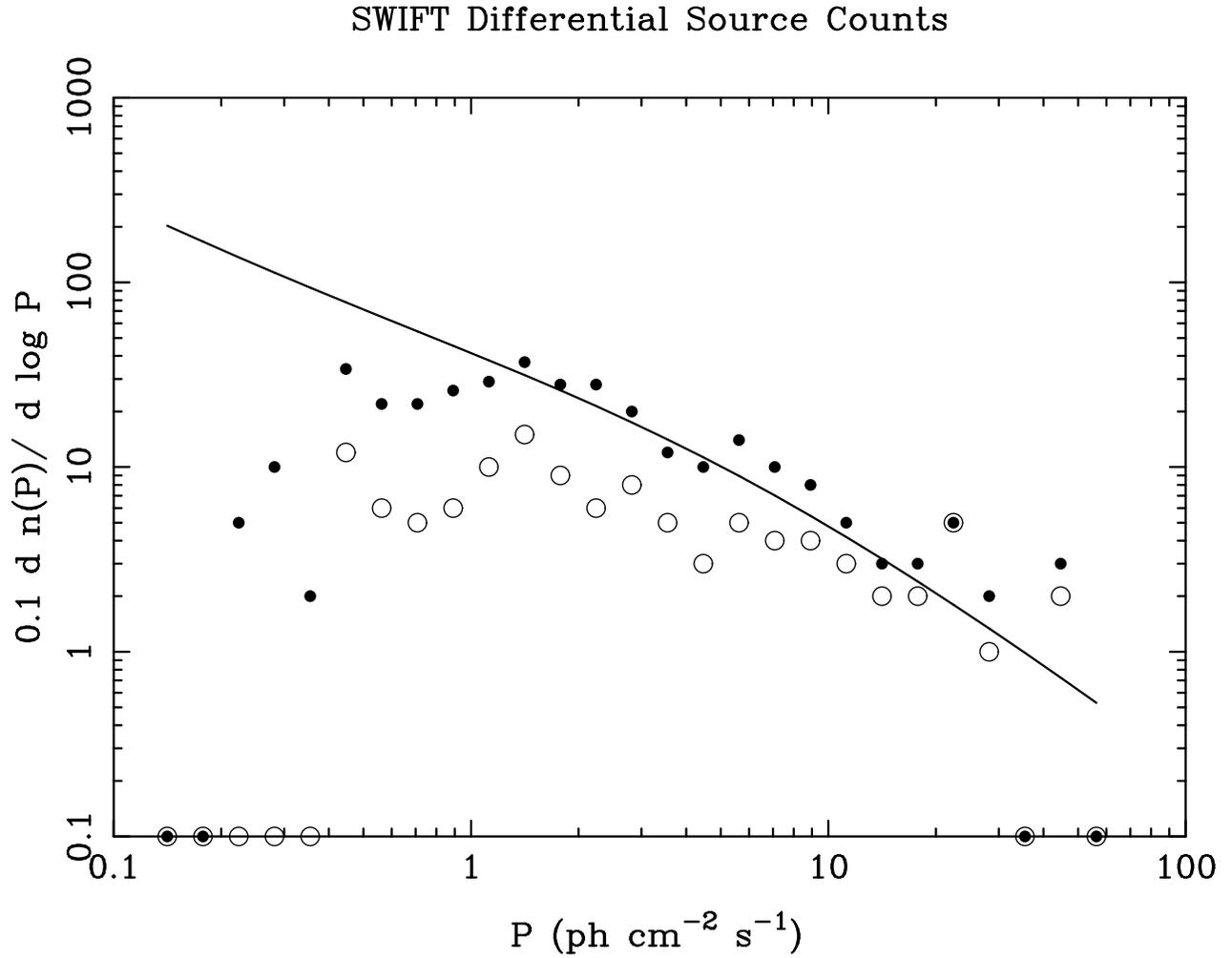}
\caption{Differential source counts for \sw: observed number (dots),
those with redshifts (circles) and the predicted number based on a 
luminosity function (line). Based on this plot, we estimate that the 
effective completeness limit for \sw\ bursts is 
$P = 1.0$ ph cm$^{-2}$ s$^{-1}$.
\label{fig2}}
\end{figure}

\begin{figure}
\includegraphics[angle=270,scale=.80]{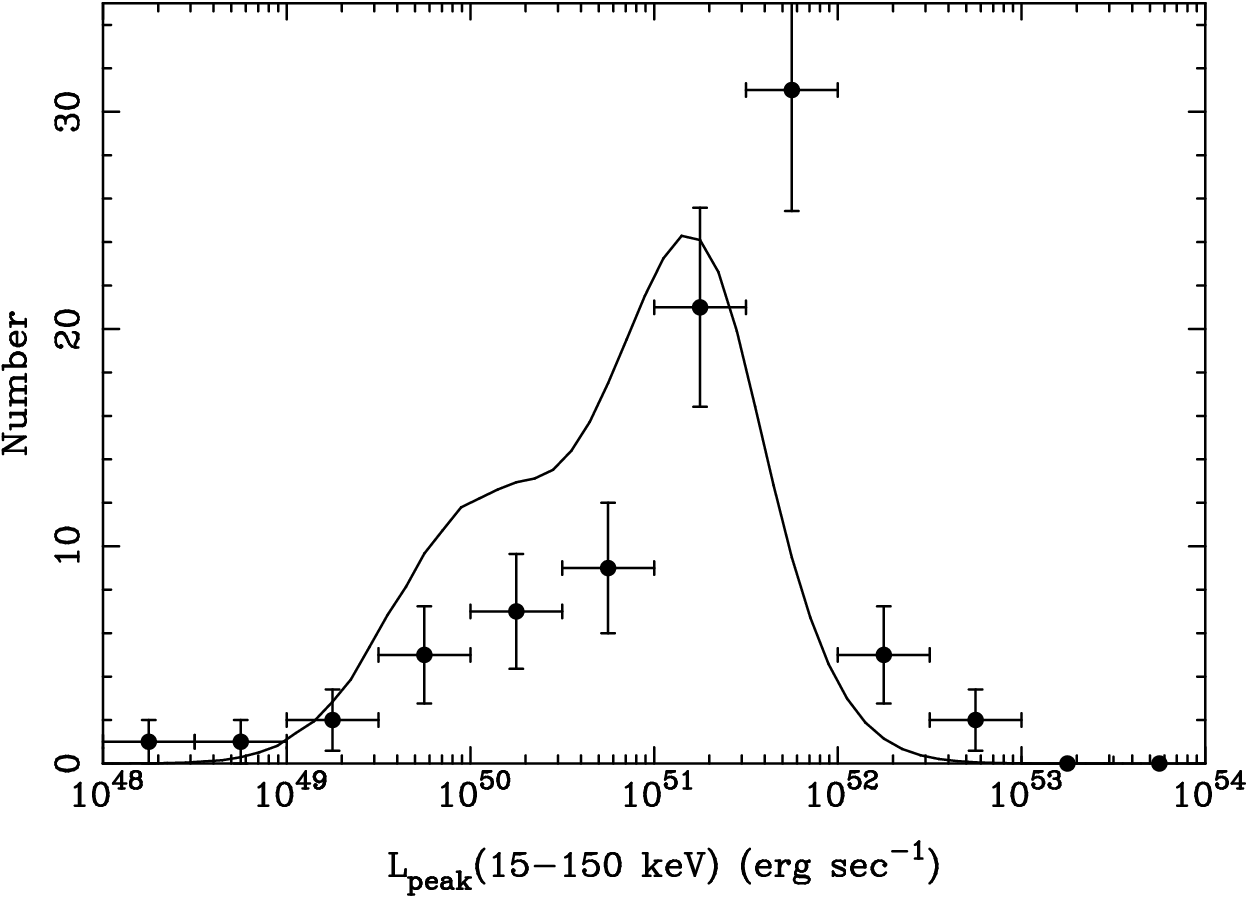}
\caption{Luminosity distribution of \sw\ sources: observed (dots)
and predicted (curve) if \gz\ equals SFH, the typical galaxy star 
formation rate, see .eq. 6.\label{fig3}}
\end{figure}

\begin{figure}
\includegraphics[angle=270,scale=.80]{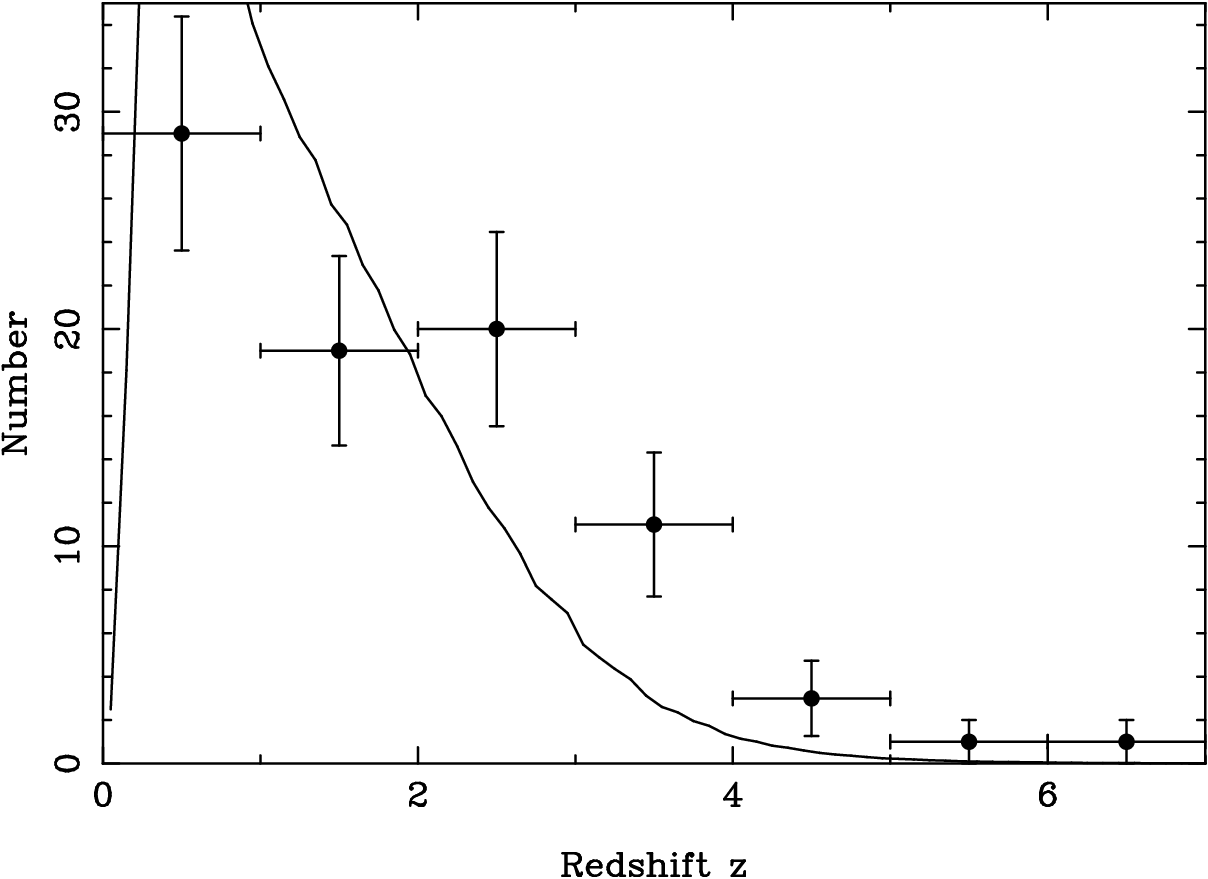}
\caption{Redshift distribution of \sw\ sources: observed (dots)
and predicted (curve) if \gz\ equals SFH, the typical galaxy star 
formation rate, see .eq. 6.\label{fig4}}
\end{figure}

\begin{figure}
\includegraphics[angle=270,scale=.80]{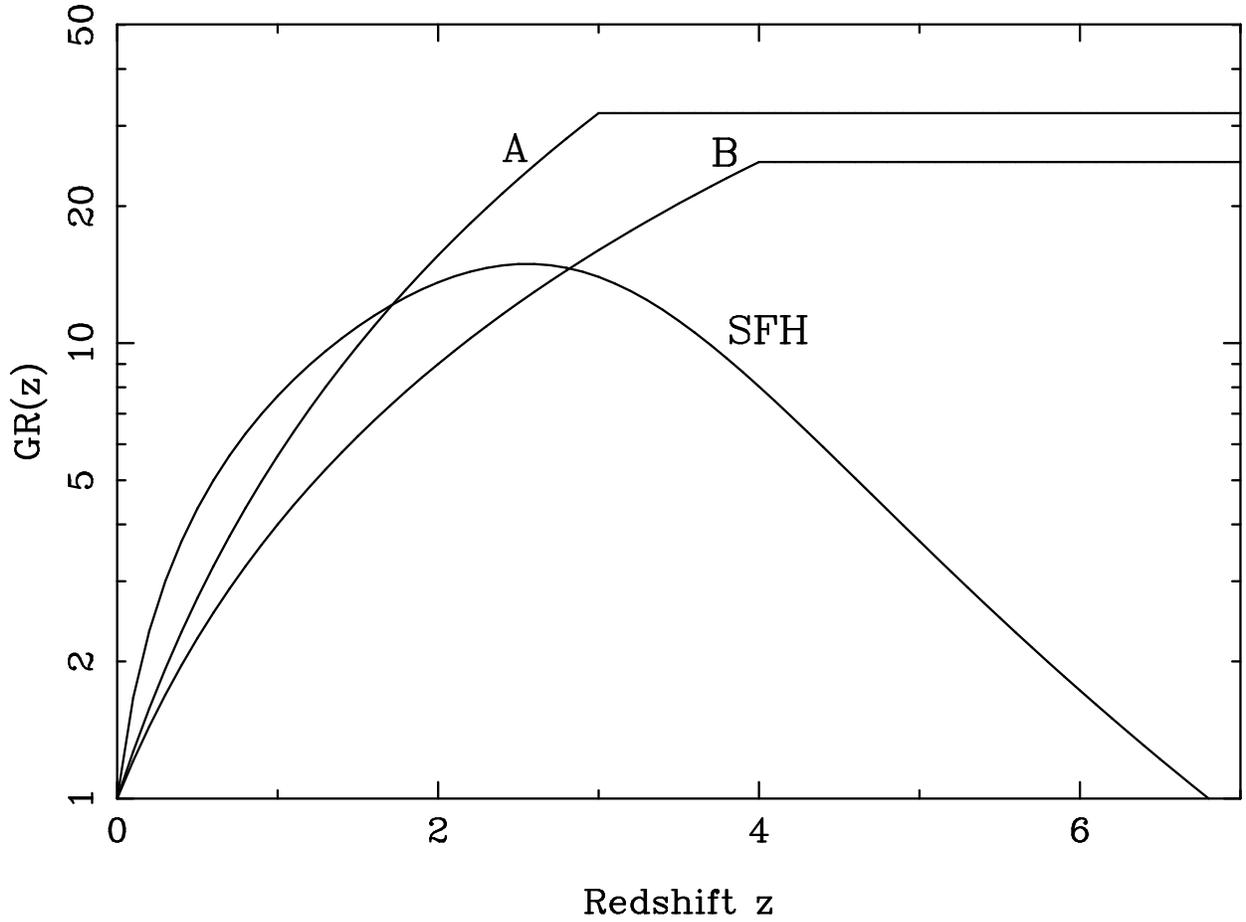}
\caption{Normalized GRB rates \gz. The typical star formation rate
SFH produces too few redshifts larger than 2. Models A and B
predict luminosity and redshift distributions that are compatible 
with the \sw\ observations.\label{fig5}}
\end{figure}

\begin{figure}
\includegraphics[angle=270,scale=.80]{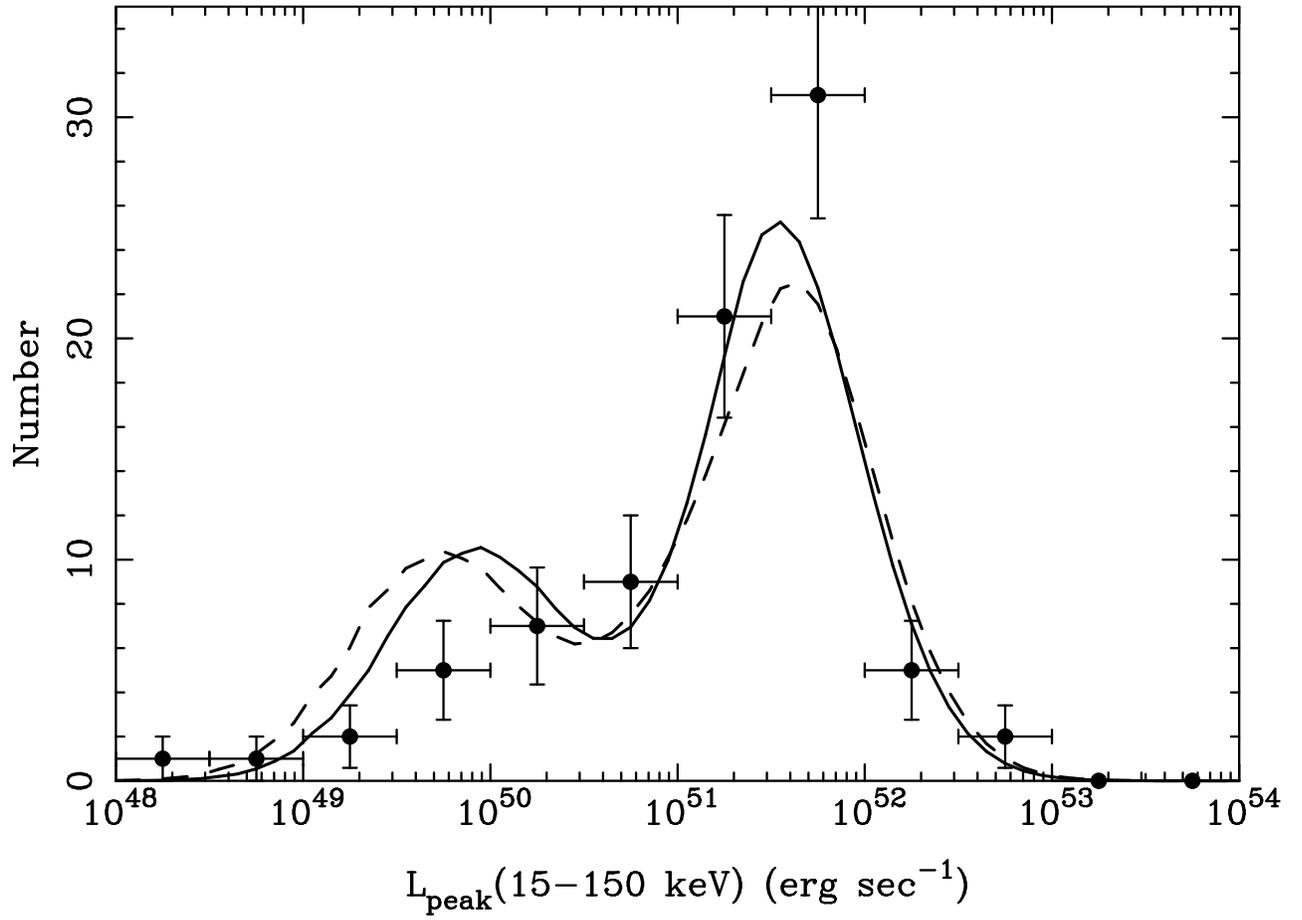}
\caption{Luminosity distribution of \sw\ sources: observed (dots)
and predicted for models A (full line) and B (dashed line).\label{fig6}}
\end{figure}

\begin{figure}
\includegraphics[angle=270,scale=.80]{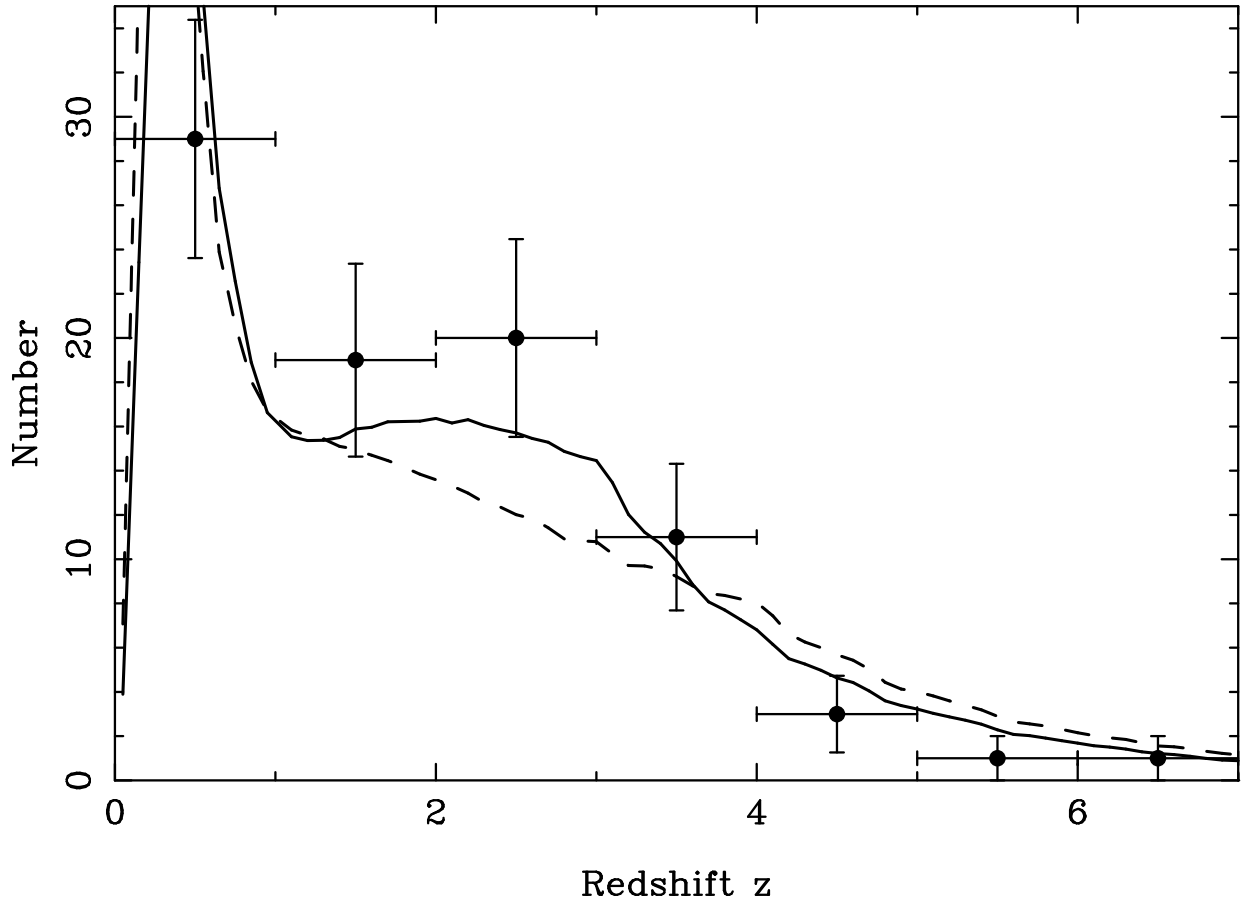}
\caption{Redshift distribution of \sw\ sources: observed (dots)
and predicted for models A (full line) and B (dashed line).\label{fig7}}
\end{figure}

\begin{figure}
\includegraphics[angle=270,scale=.80]{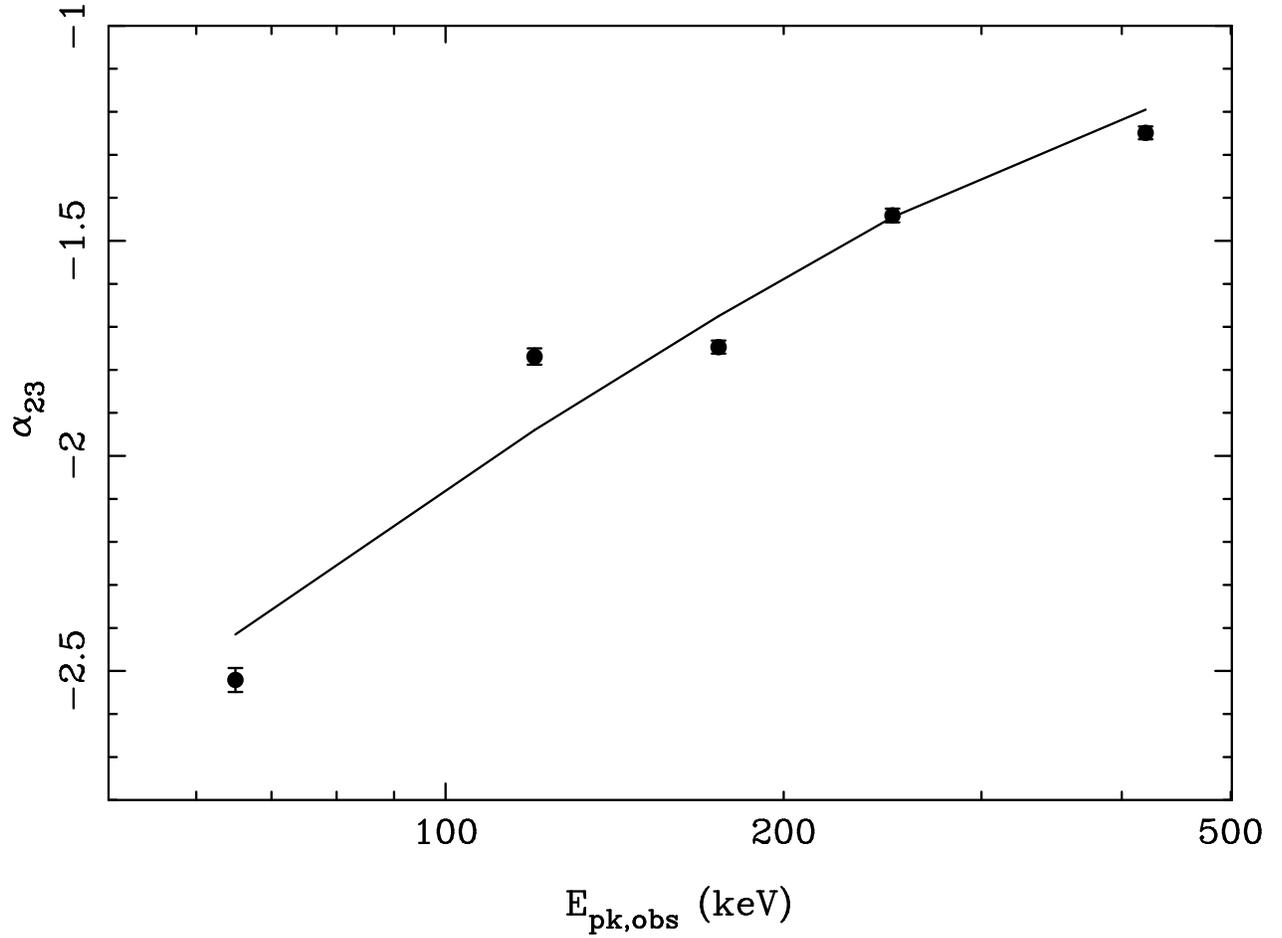}
\caption{Observed value of the photon index $\alpha_{23}$ versus 
the peak spectral energy \epob\ (points) and average values derived 
for models A and B (curve) with $\alpha = -0.8$ and $\beta = -2.6$. 
The error bars for most of the points are too small to show.\label{fig8}}
\end{figure}

\begin{figure}
\includegraphics[angle=270,scale=.80]{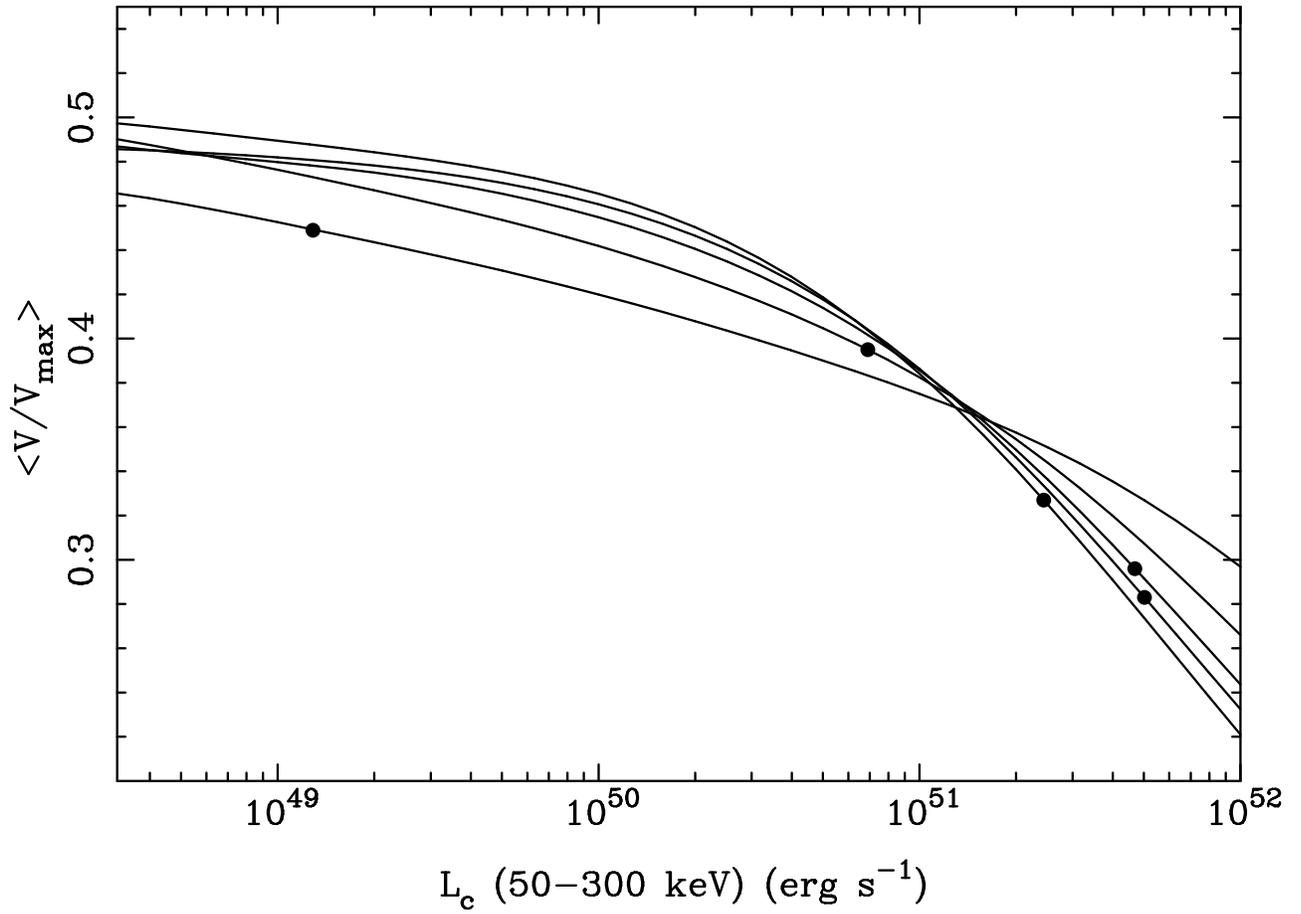}
\caption{Illustrating the derivation of $L_c(sp)$ from the observed
values of \vvmav. The points correspond to model A, see Table 4.
\label{fig9}}
\end{figure}

\begin{figure}
\includegraphics[angle=270,scale=.80]{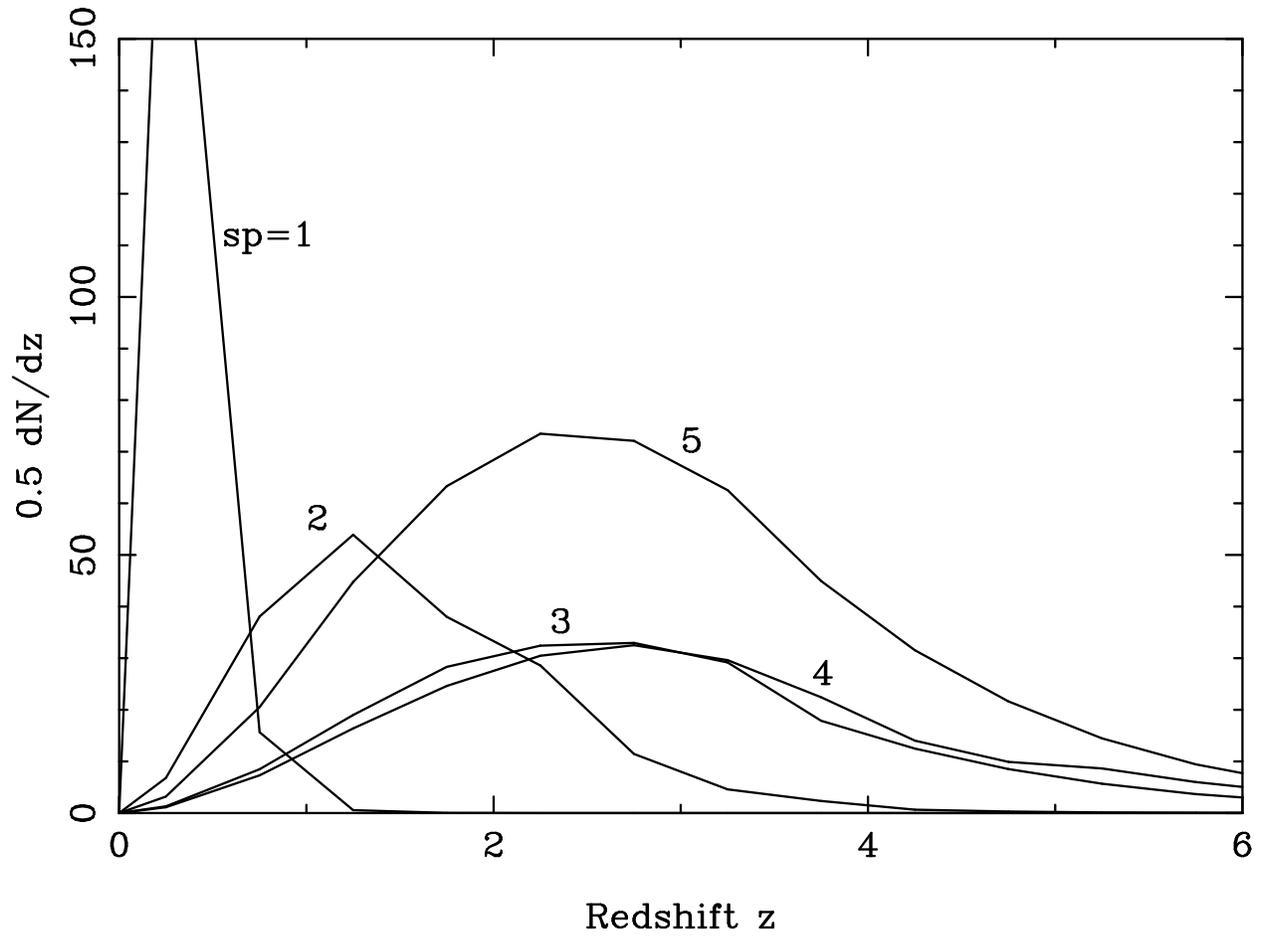}
\caption{The redshift distribution for the GUSBAD sources, for each 
of the five spectral classes, based on the \lufu\ for model A.
\label{fig10}}
\end{figure}

\begin{figure}
\includegraphics[angle=270,scale=.80]{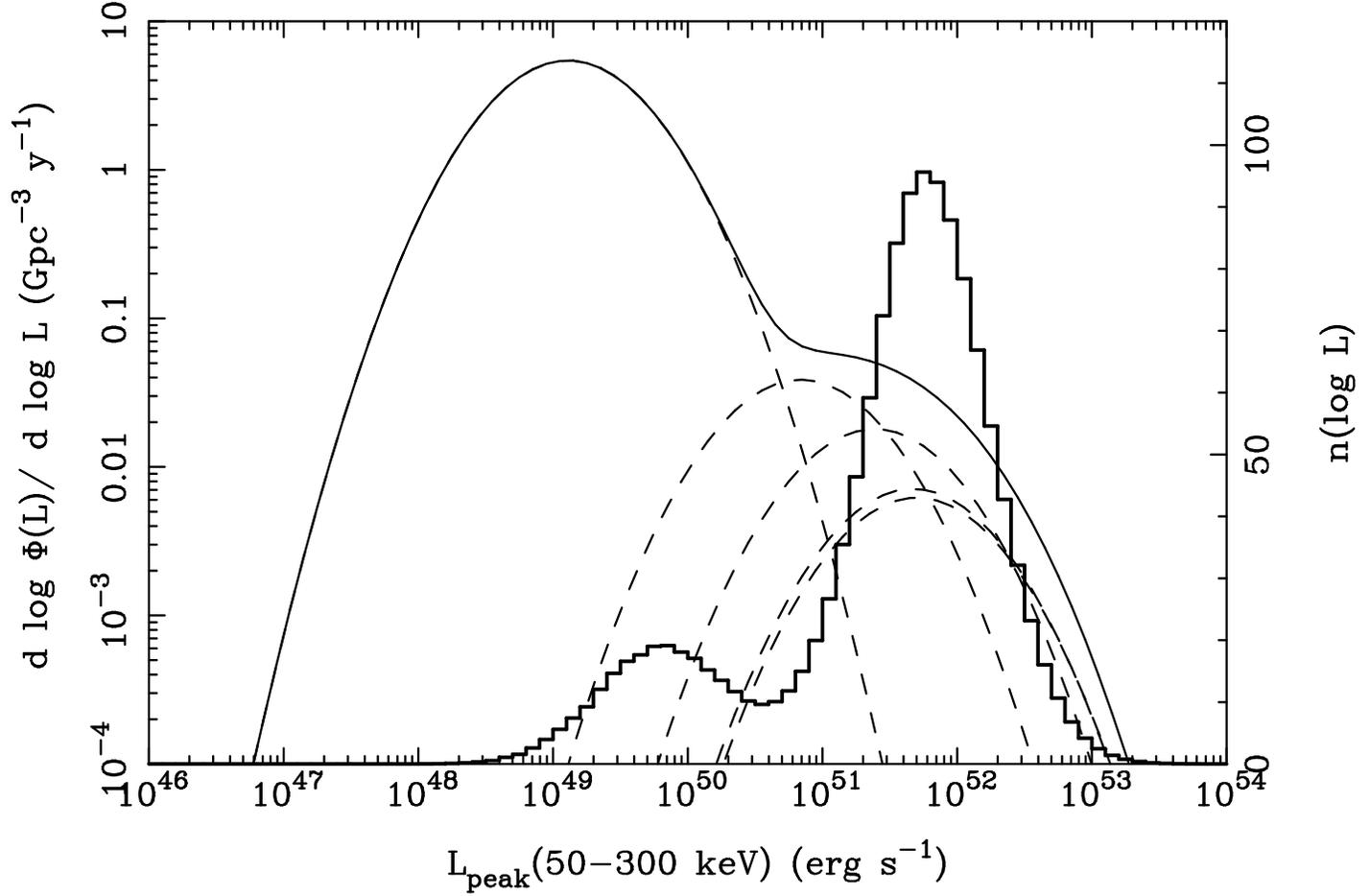}
\caption{Left side: the \lufu\ for model A (full line) and the five 
spectral class luminosity functions (dashed lines). Right side: the
corresponding luminosity distribution $n(\log L)$ in steps of 
$\Delta \log L = 0.1$ for 1319 GUSBAD sources with 
$P > 0.5$ ph cm$^{-2}$ s$^{-1}$.\label{fig11}}
\end{figure}

\begin{figure}
\includegraphics[angle=270,scale=.80]{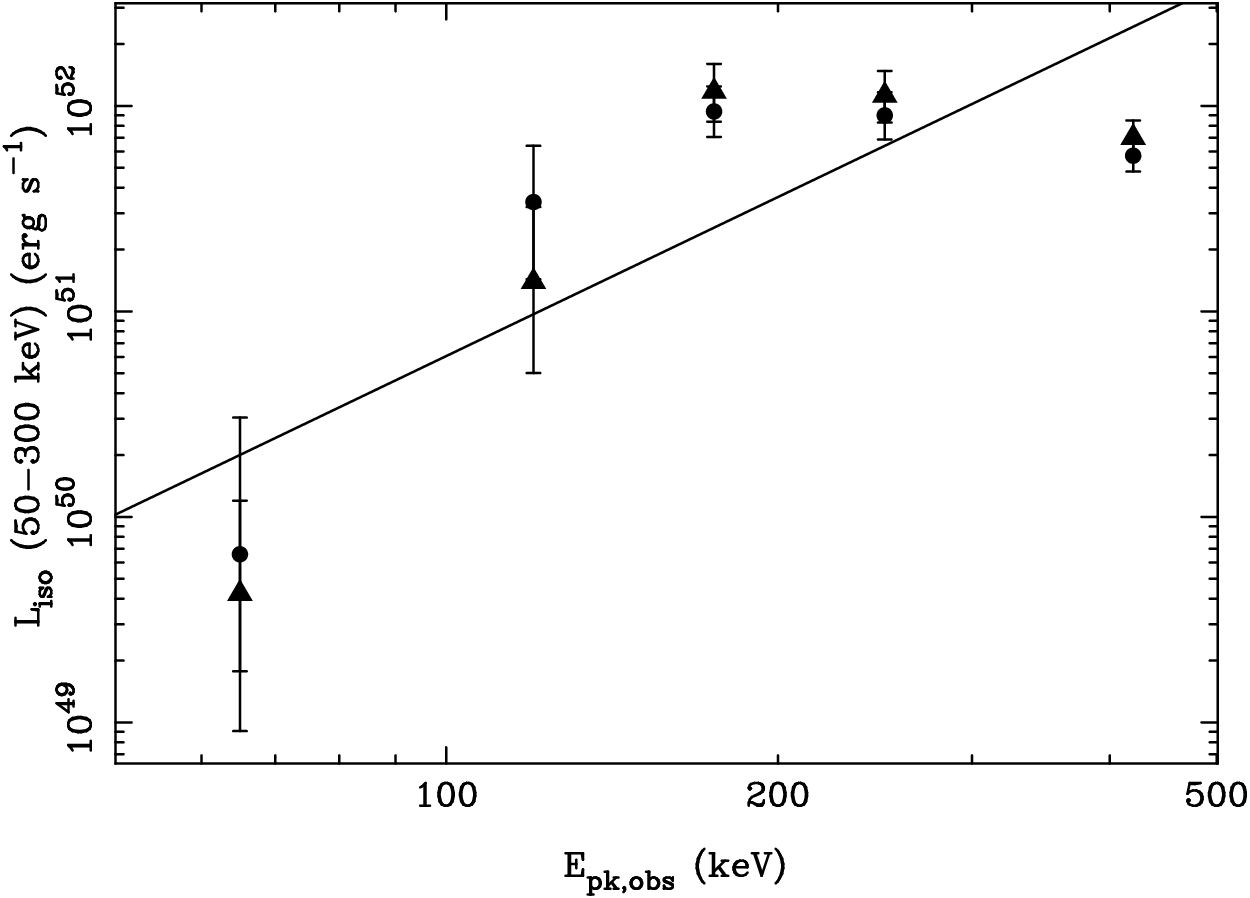}
\caption{The isotropic-equivalent luminosity \liso\ versus the 
observed \epob\ for the five spectral classes, 
for models A (circles) and B (triangles).\label{fig12}}
\end{figure}

\begin{figure}
\includegraphics[angle=270,scale=.80]{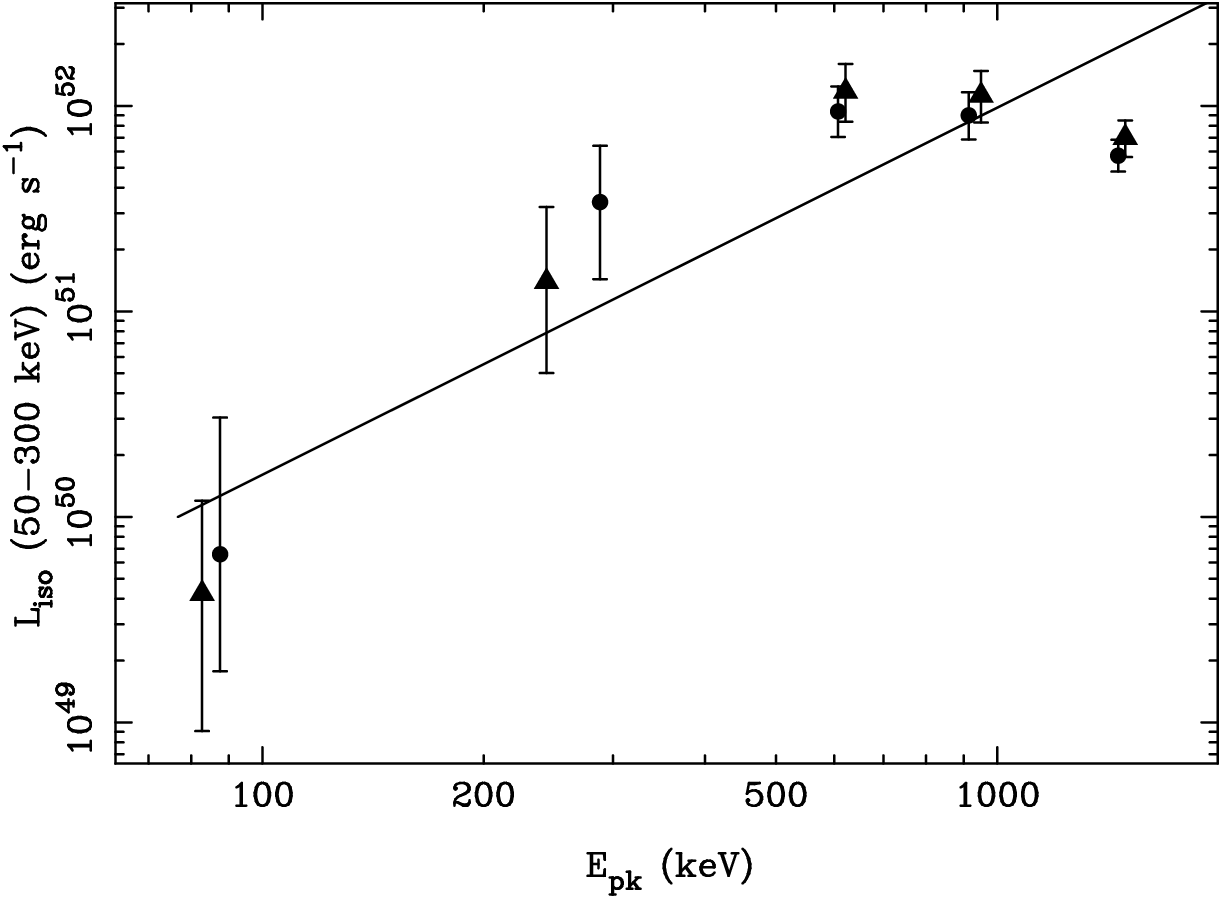}
\caption{The isotropic-equivalent luminosity \liso\ versus the 
rest frame \ep\ for the five spectral classes, 
for models A (circles) and B (triangles).\label{fig13}}
\end{figure}

\end{document}